\documentclass[preprintnumbers,onecolumn,amsmath,amssymb]{revtex4}
\usepackage{graphicx}
\usepackage{dcolumn}
\usepackage{bm}
\usepackage{epsfig}

\begin{document}

\title{Simple Model of Propagating Flame Pulsations}
\author{S. I. Glazyrin and P. V. Sasorov}
\affiliation{Institute for Theoretical and Experimental Physics, Moscow,
117218, Russia}

\begin{abstract}
A simple model which exhibits dynamical flame properties in 1D
is presented. It is investigated analytically and numerically.
The results are applicable to problems of flame propagation in supernovae Ia.
\end{abstract}

\maketitle

Observations of distant supernovae Ia explosions provide us with a very important information for 
modern cosmology (\cite{cosm1,cosm2,cosm3}). One probable scenario of type Ia supernovae is the
following: the explosions are caused by a sufficiently fast
thermonuclear burning of white dwarfs close to the Chandrasekhar limit (\cite{SNIa}).
After many years of atrophysical research, the accepted picture is that a supernova Ia explosion
usually starts as a flame front, propagating slowly due to the thermal conduction (subsonic deflagration).
Then the flame is transformed, due to a yet unknown reason, into a supersonic
detonation wave (\cite{HillebrandtNiemeyer2000}).
The main observational properties of the supernovae Ia explosions must depend significantly on the 
flame propagation process, on the transition to detonation and on the instabilities which are inherent to 
both slow and fast propagation regimes.
These are presently far from being satisfactorily understood. They pose
a challenge
for theoretical physics, hydrodynamics and mechanics, because the supernovae Ia explosions cannot be
investigated experimentally, and interiors of white dwarfs are not directly observable.

In this paper we present and investigate, analytically and by means of numerical simulations,
a very simple model of pulsating instability of a subsonic deflagration under conditions
typical for white dwarfs. The instability occurs already in
a one-dimensional geometry,
so our model is one-dimensional. Because of the strong quantum degeneracy of the electron gas in 
white dwarfs, there is only
a very weak expansion of the two-component electron-nuclei gas at the deflagration front.
Combined with the one-dimensional geometry, this means 
that the motion of the fuel plays only a  minor role, so we will consider the fuel as motionless. 
The reacting species in white dwarfs are nuclei,
whereas heat is transported by relativistic electrons and photons. This means that the 
Lewis number {\sl Le}, characterizing the relative role of
the thermal conduction and fuel diffusion, is much higher than 1: $\mbox{\sl Le}\gg 1$, 
and that the diffusion of reactants can be neglected.
The only relevant dimensionless number, incorporated in our model, is the
so-called Zeldovich number {\sl Ze}. It characterizes the
steepness of the reaction rates dependence on the temperature, or the thickness of the conductive region of the
flame front relative to the effective thickness of a layer with reactions.
Exact definitions of the Zeldovich number
in the context of our model will be presented below. Usually the subsonic deflagration becomes 
pulsating for sufficiently high {\sl Ze}.
Our model may help to get insight into mechanisms of pulsations of the subsonic deflagration.
It may also 
provide useful tests for more or less
sophisticated numerical codes used for simulations of {\sl CO} burning in white dwarfs.
It cannot give, however, a final answer about
the existence of pulsations in real flames. Our main goal is to develop a fully analytically solvable
model of burning in order to check whether the
critical {\sl Ze} number for the instability, predicted by the theory, coincides with that obtained in a
numerical experiment. For the sake of analytical solvability we will crudely simplify some
properties of the matter in white dwarfs. Our model can be considered as a further
simplification of the model considered by \cite{Siv}.

Pulsating regimes of the subsonic deflagration have attracted considerable attention for a 
long time (\cite{BZ,W,BM}).
Our attention to this problem was inspired additionally by a long-lived
controversy between the results of  \cite{Woo} and
\cite{BychkovLiberman1995}.
In \cite{BychkovLiberman1995} it was
argued that the subsonic deflagration in white dwarfs undergoes a pulsating instability.
This instability  can have important consequences for the deflagration-to-detonation transition. 
However, numerical simulations of \cite{Woo} did not show any instability.
The difference between these two results may be caused by
different and complicated databases for nuclear reactions and/or by peculiarities of numerical methods.
Our simple model, introduced below, may help separate these two quite different sources of the controversy.

Here is a plan of the remainder of the paper. Sec.~\ref{sec:model}
is devoted to the formulation of our simple model. It also contains a traveling 
wave solution 
for arbitrary {\sl Ze}. Linear stability of the traveling wave solution is investigated
in Sec.~\ref{sec:fl_fr_st}. We find there the critical Zeldovich number $\mbox{\sl Ze}_{cr}$,
below which the traveling wave is stable against small perturbations.
Sec.~\ref{numerics} is
devoted to numerical simulations of the front propagation below and above the threshold.
The results of this section make it clear that adequate codes are needed for simulations of the 
subsonic deflagration, and for a better
understanding of nonlinear pulsating regimes of flame propagation.
Our conclusions are presented in the last section of the paper.

\section{The model}
\label{sec:model}

Our model includes two dependent variables: the temperature $T$ and the deficient reagent fraction $c$. 
There are also two independent variables: time $t$ and distance $x$. Neglecting diffusion of 
the deficient reagent (which is reasonable, because the kinetic coefficients in 
a white dwarf are such that $\mbox{\sl Le}\gg1$), we 
introduce the following system of equations for $T$ and $c$:
\begin{equation}
C\partial_t T=\kappa \partial_x^2 T+W\omega
  c\Theta(T-T_0),~~\partial_t c=-\omega c\Theta(T-T_0),
\label{Ph010}
\end{equation}
where $\Theta(\dots)$ is the step-function.  These equations
describe deflagration burning in solid propellants, because two main physical processes 
which drive a slow front are present here:
the thermal conductivity and the burning itself.
The equations include the following constants: $C$ is
the thermal capacity of the fuel per unit volume;
$W$ is energy per unit volume of the
deficient reagent fraction, released due to the reaction; and
$\omega$ is the reaction rate, with dimensions
1/s.

Assuming that the deficient reagent fraction before ignition is equal to $c_0$, one 
can see that the fuel temperature after complete burnout will 
increase by the value
\begin{equation}
T_f=\frac{Wc_0}{C}\, .
\label{Ph020}
\end{equation}
Assuming further that the temperature $T(0)$ before 
ignition is much lower than $T_f$ we will neglect the former one, setting $T(0)=0$.

In real white dwarfs, the
thermal capacity depends on the temperature: for high density $\rho\sim
10^8\div 10^9$ g/cm$^3$ all parameters are determined by the relativistic degenerate electrons, that
is $C\propto T$. But here we omit the dependence for the sake of
possibility of analytical analysis. The latter concerns also the coefficient of 
thermal conduction $\kappa$ which is determined by both electrons and photons.

\begin{figure}
\includegraphics[width=3 in, clip=]{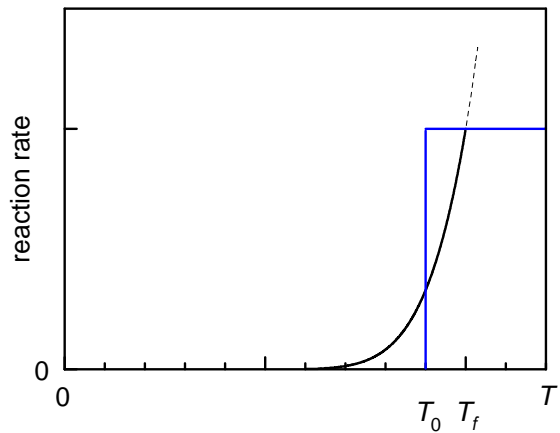}
\caption{Comparison of Arrhenius law for reaction rate and our step wise function for the same
{\sl Ze}~$=9$.}
\label{Ze-fig}
\end{figure}

The step wise approximation of the temperature dependence of nuclear reactions rates are used
to have linear equations in both domains
$T<T_0$ and $T>T_0$. Intrinsic nonlinearity of the problem is
moved only to boundary conditions between the two regions.
This simplification will allow us to investigate the problem analytically.
However we believe that our
approximation of the temperature dependence represents to some extend
very sharp temperature dependence of nuclear reaction rates, especially for
high values of the Zeldovich number: {\sl Ze}$\gg 1$. See Fig.~\ref{Ze-fig}.

Usually the Zeldovich number {\sl Ze} is defined assuming Arrhenius law for the 
temperature dependence of reactions rates: $w\propto \exp (-T_a/T)$. In this case 
the Zeldovich number is defined usually as: {\sl Ze}$=T_a/T_f$. To approximate this 
temperature dependence we may equate the temperatures for both dependence 
where the rates become $e$-fold lower than their maximum values at $T=T_f$. Then we 
will have the following definition of $T_0$ through $T_a$ and $T_f$:
\begin{equation}
{\sl Ze}=\frac{T_a}{T_f}=\frac{T_0}{T_f-T_0};\quad\mbox{~~or~~} T_0=T_f\, \frac{T_a}{T_f+T_a}=T_f\,  \frac{{\sl Ze}}{1+{\sl Ze}}
\, .
\label{Ze}
\end{equation}
Applying the same procedure to the power law for the reaction rate ($w\propto T^n$), we would obtain the following relationship:
$T_0=T_fe^{-1/n}\approx T_f\, (1-n^{-1})$ (for $n\gg 1$).

We may introduce now new dimensionless variables labelled by ``$\tilde{\phantom{a}}$'' and defined as follows:
\begin{equation}
t=\tau \tilde{t}\, ;\quad x=l\tilde{x}\, ;\quad c=c_0\tilde{c}\, ;\quad T=T_f\tilde{T}\, ,
\label{Ph030}
\end{equation}
where
\begin{equation}
l=\sqrt{\frac{\kappa}{C\omega}\,\left(\frac{T_f}{T_0}-1\right)}\, ;\quad \tau=\frac{Cl^2}{\kappa}\, .
\label{Ph040}
\end{equation}
Scales for $x$, $t$, $T$, $c$ are chosen for all characteristic parameters of the traveling wave in dimensionless units to be equal to 1. This concerns the velocity of the traveling wave, the concentration $c$ far before the wave, the temperature $T$ far beyond the wave, and the characteristic width of the wave. We will use below only the dimensionless variables skipping the tilde ``$\tilde{\phantom{a}}$''.

Thus our dimensionless system is as follows:
\begin{equation}
  \partial_t T=\partial_x^2 T+\omega_0
  c\Theta(T-T_0),~~\partial_t c=-\omega_0 c\Theta(T-T_0)\, ,
\label{sys:simpl}
\end{equation}
where
\begin{equation}
\omega_0=\omega\tau = T_0/(1-T_0); \quad (0<T_0<1)\, .
\label{T0}
\end{equation}
Dimensionless value of the igniting temperature $T_0$ is used in Eq.~(\ref{T0}). In initial physical units it is equal to $T_0/T_f$.
Our effective Zeldovich number [Eq.~(\ref{Ze})] in dimensionless units can be expressed as follows:
\begin{equation}
\mbox{\sl Ze}=\omega_0=\frac{T_0}{1-T_0}\, .
\label{Ze_m}
\end{equation}

The stationary traveling wave depends on $x$ and $t$ only via the combination $\xi=x-vt$.
For this reason it is convenient to introduce new spacial coordinates $\xi=x-vt$ instead of $x$, where $v$ is the constant velocity of moving frame.
Thus $T$ and $c$ depend now on $t$ and $\xi$.

It is important that the
system~(\ref{sys:simpl}) is linear in both $T>T_0$ and $T<T_0$. Nonlinearity appears only at the
points, where the transition $0\rightarrow 1$ in $\Theta$-function occur. Our equations~(\ref{sys:simpl}) mean that $T$, $\partial_x T$ and $c$ are continues at $T=T_0$. These conditions can be treated as matching conditions for solutions of the linear equations at $T<T_0$ and $T>T_0$.
As a result we may obtain the following system of equations instead of the system~(\ref{sys:simpl}) for monotonic in $\xi$ functions. It is
convenient for the investigation of the stationary traveling wave and its
linear stability: for $T_{+}>T_0$
\begin{equation}
  \partial_t T_{+}=v \partial_{\xi} T_{+} +\partial_{\xi}^2 T_{+}+\omega_0 c_{+},~~\partial_t c_{+}=v \partial_{\xi} c_{+}-\omega_0 c_{+},
\label{moving_1}
\end{equation}
and for $T_{-}<T_0$
\begin{equation}
  \partial_t T_{-}=v \partial_{\xi} T_{-} +\partial_{\xi}^2 T_{-},~~c_{-}=1.
\label{moving_2}
\end{equation}
 Here
\begin{equation}
  f(\xi,t)=\left\{
  \begin{array}{lcr}
  f_{-}(\xi,t)&\mbox{~~for~~}& \xi>\xi_f(t),\\
  f_{+}(\xi,t)&\mbox{~~for~~}& \xi<\xi_f(t),
  \end{array}
  \right.
\label{moving_3}
\end{equation}
where $f$ represents $T$ or $c$, and $\xi_f(t)$ is position of the front, where $T=T_0$. The matching conditions read as:
\begin{eqnarray}
T_{+}(\xi_f(t),t)&=&T_{-}(\xi_f(t),t)=T_0;\nonumber\\
\partial_\xi T_{+}(\xi_f(t),t)&=&\partial_\xi T_{-}(\xi_f(t),t);\nonumber\\
\quad c_{+}(\xi_f(t),t)&=&1.
\label{match}
\end{eqnarray}

The stationary traveling wave obeys the following boundary conditions
\begin{equation}
  \xi\rightarrow\infty:~T=0,~c=1;~~\xi\rightarrow -\infty:~\partial_x   T=0,~~c=0,
  \label{BC}
\end{equation}
and the conditions:
\begin{equation}
\partial_t c=\partial_t T=\xi_f(t)=0 \, .
\label{stat}
\end{equation}
The latter equality originates from a freedom due to the translation invariance of the problem, and hence from possibility to set the front at an arbitrary point of the $\xi$-space. 

The system~(\ref{moving_1})-(\ref{stat}) presents nonlinear eigen-value problem, with $v$ being the eigen-value. There is unique solution of this problem:
\begin{eqnarray}
&&~~v=1\nonumber\\
  &\xi>0:&~~c=1,~~T=T_{-}=T_0e^{-\xi},\nonumber\\
  &\xi<0:&~~c=c_{+}=e^{\omega_0\xi},~~T=T_{+}=1-\frac{1}{\omega_0+1}e^{\omega_0\xi}\, .
  \label{sys:simpl_solution}
\end{eqnarray}
As a result we will set $v=1$ as a constant velocity of the moving frame for the equations~(\ref{moving_1}) and~(\ref{moving_2}).

We have for the traveling burning wave velocity in the initial physical units:
$$
v=\sqrt{\frac{\kappa\omega}{C}\,\left(\frac{T_f}{T_0}-1\right)}\, .
$$

\section{Flame front stability}
\label{sec:fl_fr_st}

We investigate here linear stability of the traveling wave solution obtained in the previous section. The
latter one is  called now as an unperturbed, and will be designated by the superscript ``$^{(0)}$''. The full solution is presented in the form
\begin{equation}
T=T^{(0)}+T^{(1)},\quad c=c^{(0)}+c^{(1)},\quad \xi_f=\xi_f^{(1)}\, ,
\label{Lin1}
\end{equation}
where $T^{(1)}\ll T^{(0)}$, $c^{(1)}\ll c^{(0)}$ and $\xi_f^{(1)}\ll 1$. In this case our system can be linearized. Since the equations~(\ref{moving_1}) and~(\ref{moving_2}) are linear from the very beginning, their linearization leads only to setting of the superscript ``(1)'', and to setting $v=1$. However linearization of the matching conditions is not so trivial. They are transformed to:
\begin{eqnarray}
&&\xi_f^{(1)}\partial_\xi T_{+}^{(0)}(0,t)+T_{+}^{(1)}(0,t)=0\, ,
\label{Lin2}\\
&&\xi_f^{(1)}\partial_\xi T_{-}^{(0)}(0,t)+T_{-}^{(1)}(0,t)=0\, ,
\label{Lin2a}\\
&&\xi_f^{(1)}\partial_\xi^2 T_{+}^{(0)}(0,t)+\partial_\xi T_{+}^{(1)}(0,t)\nonumber\\
&=&\xi_f^{(1)}\partial_\xi^2 T_{-}^{(0)}(0,t)+\partial_\xi T_{-}^{(1)}(0,t),
\label{Lin3}\\
&&\xi_f^{(1)}\partial_\xi c_{+}^{(0)}(0,t)+c_{+}^{(1)}(0,t)=0\, ,
\label{Lin4}
\end{eqnarray}
The boundary conditions~(\ref{BC}) give vanishing boundary conditions for $T_{+}^{(1)}$ and $c_{+}^{(1)}$ at $\xi\to -\infty$ and for $T_{-}^{(1)}$ at
$\xi\to +\infty$. There is no source for the perturbation of $c$ before the front, so we set $c_{-}=0$.

Since Eqs.~(\ref{moving_1}) and~(\ref{moving_2}) and the matching conditions~(\ref{Lin2})-(\ref{Lin4}) together with the boundary conditions at $\xi\to\pm\infty$ do not contain explicit dependence on time, their solution should have the following dependence on $t$: $\propto e^{pt}$, where $p$ is an eigen-value of corresponding eigen-value linear boundary problem. Perturbation of the front position has the similar form. Thus
\begin{equation}
  \xi_{f}(t)=\chi e^{pt}\, .
\label{fr}
\end{equation}
Eqs.~(\ref{moving_1}) and~(\ref{moving_2}) do not contain even explicit dependence on $\xi$. This property
leads to a very simple form of the solutions at $\xi>0$ and $\xi<0$:
\begin{eqnarray}
T_{-}^{(1)}&=&\alpha e^{pt+\lambda \xi}\, ,
\label{sol1}\\
c_{+}^{(1)}&=&\beta e^{pt+(p+\omega_0) \xi}\, ,
\label{sol2}\\
T_{+}^{(1)}&=&\gamma e^{pt-(1+\lambda)\xi}-\frac{\omega_0}{(p+\omega_0)^2+\omega_0}\beta e^{pt+(p+\omega_0) \xi}\, ,
\label{sol3}
\end{eqnarray}
where
\begin{equation}
p=\lambda^2+\lambda\, ,
\label{sol4}
\end{equation}
and necessary condition ${\rm Re}\lambda<0$ to obey the boundary conditions at $+\infty$.

The presentation of solution~(\ref{fr})-(\ref{sol3}) contains 4 arbitrary constants $\chi$, $\alpha$, $\beta$ and $\gamma$. However the solution should obey the matching conditions~(\ref{Lin2})-(\ref{Lin4}). As a result we have homogeneous system of 4 linear equations for 4 variables. For the system of equations have nontrivial solution its determinant should be equal to 0. As a result we obtain the following equation for the eigen value $p$
expressed trough $\lambda$ in accordance to Eq.~(\ref{sol4}) under condition that $(p+\omega_0)^2+\omega_0\neq 0$:
\begin{equation}
\frac{2\,\omega_{0}\,\lambda^3+\left(\omega_{0}^2+2\,\omega_{0} \right)\,\lambda^2+
2\,\omega_{0}^2\,\lambda+\omega_{0}^2}{\left(\omega_{0}+1\right)\,\lambda^2+\omega_{0}^2+\omega_{0}}
=0\, .
\label{det}
\end{equation}

\begin{figure}
 \includegraphics[height=0.45\textwidth,angle=270]{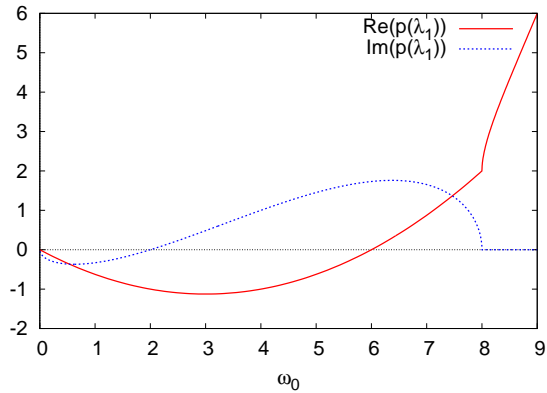}
 \caption{\label{fig:p1} The complex growth rate $p_1$ versus $\omega_0=\mbox{\sl Ze}-1$.}
\end{figure}

\begin{figure}
 \includegraphics[height=0.45\textwidth,angle=270]{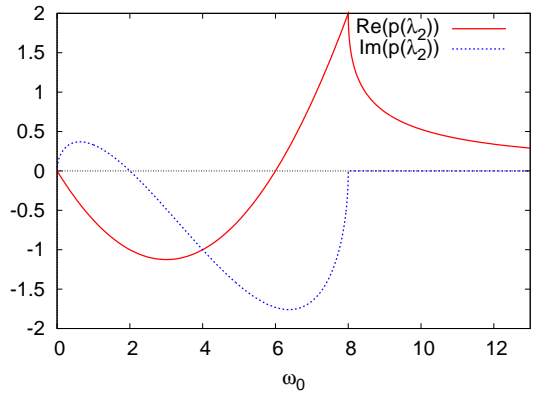}
 \caption{\label{fig:p2} The complex growth rate $p_2$ versus $\omega_0=\mbox{\sl Ze}-1$.}
\end{figure}

The cubic polynomial entering into the nominator of Eq.~(\ref{det}) has a root corresponding to $p=0$ that could be assumed beforehand (see below). As a result solution of Eq.~(\ref{det}) is reduced to quadratic equation. Thus we have the following set of eigen values in terms of $\lambda$:
\begin{eqnarray}
  \lambda_1&=&-\frac{\sqrt{\omega_{0}^2-8\,\omega_{0}}+\omega_{
 0}}{4},\nonumber\\
\lambda_2&=&\frac{\sqrt{\omega_{0}^2-8\,\omega
 _{0}}-\omega_{0}}{4},\nonumber\\
\lambda_3&=&-1.
\end{eqnarray}
The 3rd root $\lambda_3=-1$ gives $p_3=0$, $\alpha_3=-\omega_0/(1+\omega_0)$, $\beta_3=1$, $\gamma_3=0$. This eigen solution corresponds in according to Eq.~(\ref{sys:simpl_solution}) to the small shift of the stationary traveling front as a whole. Existence of such solution could be supposed in advance because of a translational symmetry of the problem. Thus physical roots are $\lambda_{1,\, 2}$. Real and imaginary parts of
$p_{1,\, 2}$ versus $\omega_0$ are plotted in Figs.~\ref{fig:p1} and~\ref{fig:p2}.  If $\omega_0\geq8$, so that both $\lambda_1$ and $\lambda_2$ are real, then both eigen values $p_{1,\, 2}$ are positive. If, however, $\omega_0<8$, then
$$ 
p_{1,\, 2}=\frac{\omega_0^2-6\omega_0\pm i\left(\omega_0-2\right)\sqrt{8\omega_0-\omega_0^2}}{8}\, .
$$
Hence the traveling wave solution is stable against small perturbations at $\omega_0<6$ and unstable in the opposite case $\omega_0>6$. The eigen values at the threshold $\omega_0=6$ are equal to $p_{1,\, 2}=\pm 4 i /\sqrt{3}$. Thus the perturbations at the threshold become purely oscillating.

Expressing this result in terms of effective Zeldovich number {\sl Ze}, introduced above, we may say that subsonic deflagration, having the form of the traveling wave, is stable in the frame of our model against small perturbations at $\mbox{\sl Ze}<\mbox{\sl Ze}_{cr}=6$. There is no stable stationary moving flame front at $\mbox{\sl Ze}>\mbox{\sl Ze}_{cr}=6$ in the frame of our model. Since time dependence of perturbations at the threshold $\mbox{\sl Ze}=\mbox{\sl Ze}_{cr}$ are purely oscillating, one may assume that the subsonic deflagration at
$\mbox{\sl Ze}>\mbox{\sl Ze}_{cr}=6$ becomes oscillating.


\section{Numerical results}
\label{numerics}

\subsection{The original model}
\label{sec:num_res}

Such a simple system can be calculated numerically and it is a good
test for analytical predictions. We use
the Crank-Nicolson finite-difference method to solve
the system~(\ref{sys:simpl}). Initial and boundary conditions are set in
two ways. In the first a task of self-formation of the flame by a hot wall
is set. In the whole region of calculation $[0;L]$ cold unburned
matter is put, but the left wall is hot:
\begin{eqnarray}
  T(t,x=0)=1,~~T(t,x=L)=0,\nonumber\\
  T(t=0,x\in (0,L))=0,~~c(t=0,x\in (0,L))=1.
\end{eqnarray}
If in this case a stationary front appears then it is natural
for the system.
The second way is to set distribution of the temperature and the concentration
according to the analytical solution (\ref{sys:simpl_solution}), and
to observe its evolution.

Numerical methods suppose discritization of space $dx$ and time
$dt$. To obtain the physically correct solution we should properly choose
these quantities. The first condition is $dt=10^{-2}/\omega_0$. This implies
that only
1\% of matter will burn on every numerical step, and therefore prohibits
abrupt changes in the solution, and in such a way controls
fluctuations.
The second condition comes from the analytical solution. It follows from Eq.~(\ref{sys:simpl_solution})
that there are two typical lengths in the system: 1 and $1/\omega_0$. To resolve
every change we need $dx\ll dx_0=\min (1, 1/\omega_0)$. During numerical
simulation several calculations were made with $dx>dx_0$ to determine
a consequence of wrong discretization.

\begin{figure}
 \includegraphics[height=0.45\textwidth,angle=270]{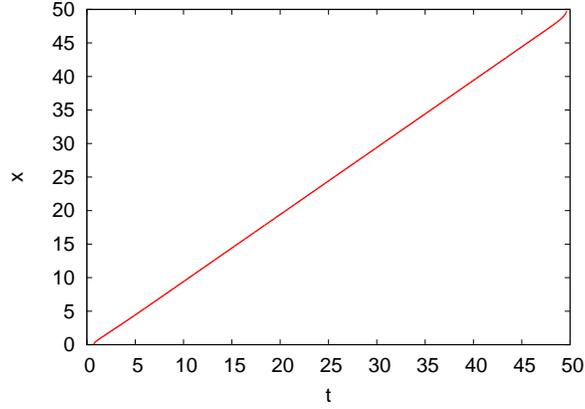}
 \caption{\label{fig:omega0_1} The front position $x(t)$ versus time $t$ for $\omega_0=1$.}
\end{figure}

The front coordinate $x$ is determined by point $c=0.5$ (the definition
diverge with the theoretical definition $T=T_0$, but in the case of the stationary wave the
points will be located on a constant distance from each other). The dependence of $x(t)$ for the simulation with
$\omega_0=1$ ($dx\ll dx_0$) is shown in
Fig.~\ref{fig:omega0_1}. Good linear curve means that
the front moves with a constant velocity and is stable. By fitting
$dx/dt$ the front velocity could be found and it is $v=1.00$, what is a very good agreement
with the theory.

\begin{figure}
 \includegraphics[height=0.45\textwidth,angle=270]{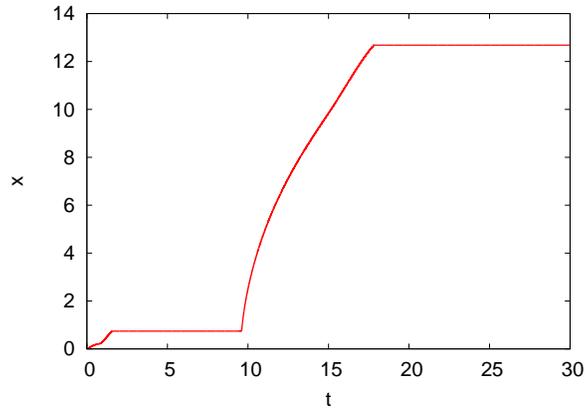}
 \caption{\label{fig:omega0_7} The front position $x(t)$ versus time $t$ for $\omega_0=7$.}
\end{figure}

But it must be noted that despite good linear dependence, 
observed on large
scales in Fig. \ref{fig:omega0_1}, small scales exhibit pulsations. These pulsations are fully
a numerical effect, because their period is exactly $T=dx/v$ (it can be
easily tested by simulations with different $dx$).

The evolution of front position $x(t)$ when $\omega_0=7$ is
shown in Fig. \ref{fig:omega0_7}. According to the theory such regime should be unstable, what occurs
in the simulation: a regime of moving with nonconstant velocity
interchanges with the regime of the front standing. For the
model~(\ref{sys:simpl}), the front will not move after $t=20$. This fact is
interpreted below.

\begin{figure}
 \includegraphics[height=0.5\textwidth,angle=270]{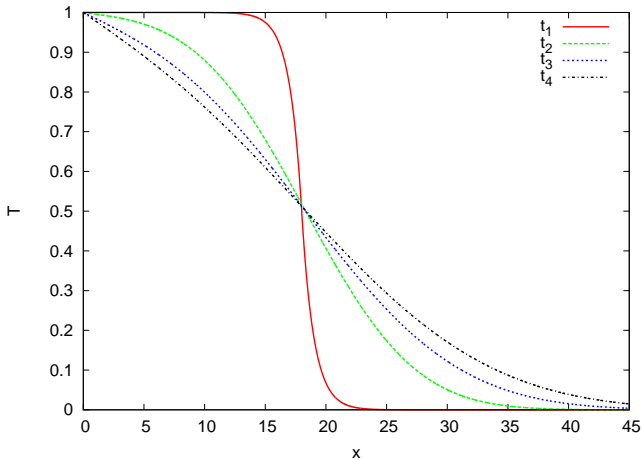}
 \caption{\label{fig:Tt_80_1} Front profiles in terms of $T$ versus $x$ for $\omega_0=6$ ($T_0\approx0.857$) at
  different time moments: $t_0<t_1<t_2<t_3$.}
\end{figure}

Table \ref{tab:toy1} presents results for the first way of initial and
boundary conditions set. The commentary column in the table shows propagation
regime: ``flame'' means the flame propagation, ``therm'' is the evolution of
medium parameters like in Fig.~\ref{fig:Tt_80_1} (which corresponds to
the unstable regime). In this case
the temperature undergo evolution like thermoconductivity without burning.
Combinations of the terms ``flame'' and ``term'' in Table~\ref{tab:toy1}
mean that  there are transition stages in these cases before establishing of an ultimate regime. The latter one
is the last word in the combinations.
Results for the analytical solution as an initial condition are shown in Table
\ref{tab:toy2}.
From those two tables we see that  $\omega_0=6$ is a critical point for system
parameters. When $\omega_0<6$ a stationary front in a system can
exist. When $\omega_0>6$ the front appears, but have nonconstant velocity
and live for a limited period of time. The dependence $x(t)$ for such front
(in case of first boundary condition) is shown in Fig.~\ref{fig:omega0_7}.

\begin{table}
\caption{\label{tab:toy1} Results for wall ignition:}
\begin{center}
\begin{tabular}{|c|c|c|c|}
  \hline
  $\omega_0$ & $dx$ & $v$ & comm. \\
  \hline
  1.0 & 0.05 & 1.000 & flame \\
  4.0 & 0.05 & 0.996 & flame \\
  5.0 & 0.05 & 0.992 & flame \\
  5.5 & 0.05 & 0.993 & therm-flame \\
  5.8 & 0.05 & 0.993 & therm-flame \\
  6.0 & 0.05 & 6.15$\div$1.24 & therm-flame \\
  6.5 & 0.05 & 5.56$\div$0.99 & therm-flame-therm \\
  7.0 & 0.05 & 6.40$\div$1.04 & therm-flame-therm \\
  8.0 & 0.05 & 4.32$\div$1.08 & therm-flame-therm \\
  \hline
  1.0 & 1.5 & 0.711 & flame \\
  4.0 & 0.5 & 4.74$\div$0.94 & therm-flame-therm \\
  \hline
\end{tabular}
\end{center}
\end{table}

\begin{table}
\caption{\label{tab:toy2} Analytic initial conditions:}
\begin{center}
\begin{tabular}{|c|c|c|c|}
  \hline
  $\omega_0$ & $dx$ & $v$ & comm. \\
  \hline
  1.0 & 0.05 & 1.000 & flame \\
  4.0 & 0.05 & 0.996 & flame \\
  5.5 & 0.02 & 1.006 & flame \\
  5.8 & 0.02 & 1.010 & flame \\
  6.0 & 0.01 & 1.019 & flame \\
  7.0 & 0.01 & -- & therm \\
  8.0 & 0.01 & -- & therm \\
  9.0 & 0.01 & -- & therm \\
  \hline
  1.0 & 1.5 & 0.711 & flame \\
  4.0 & 0.5 & -- & therm \\
  \hline
\end{tabular}
\end{center}
\end{table}

Here we should emphasize simulations when $dx>dx_0$ is set. Two
runs ($\omega_0=1$, $dx=1.5$ and $\omega_0=4$, $dx=0.5$) show
that a wrong $dx$ results in a wrong behavior, such as an incorrect front velocity
or an instability-like regime.

\subsection{Modification of the simple model}
\label{sec:modif}

\begin{figure}
 \includegraphics[height=0.5\textwidth,angle=270]{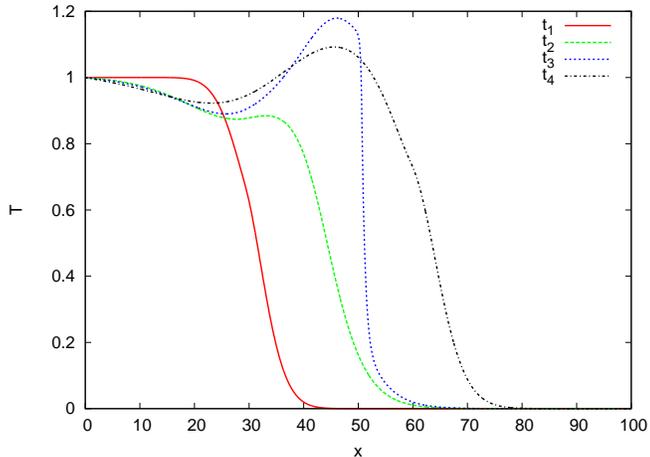}
 \caption{\label{fig:Tt_80} Sequential front profiles in terms of $T$ versus $x$ at time moments: $t_1<t_2<t_3<t_4$.}
\end{figure}

\begin{figure}
 \includegraphics[height=0.5\textwidth,angle=270]{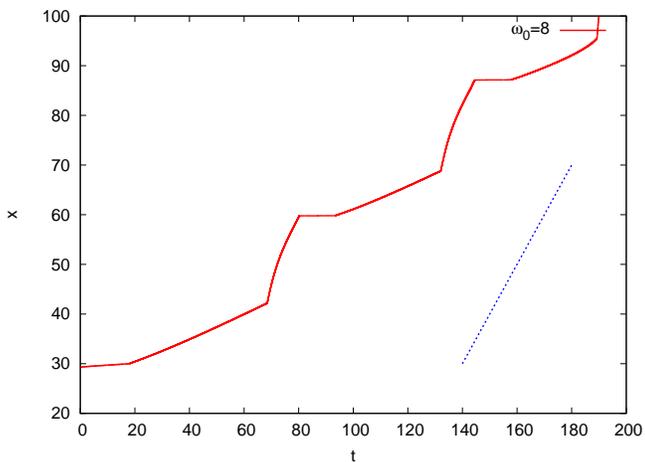}
 \caption{\label{fig:omega8_mod} Front propagation in terms of $x(t)$ for $\omega_0=8$ in modified
   model (solid line). For comparison dotted line displays moving with constant velocity
   $v=1$.}
\end{figure}

The model suffers from some nonphysical effects: when $T<T_0$
the burning rate drops to zero. This is why the flame stops at certain moment
of time in unstable regime and never runs again (only heating by the left wall could
ignite further burning). Lets consider a small modification of the burning rate:
\begin{equation}
  \partial_t T=\partial_x^2 T+R(c, T),~~\partial_t c=-R(c, T),
  \label{sys:modif}
\end{equation}
\begin{equation}
  R(c,T)=\omega_0c\Theta(T-T_0)+\omega_1c\frac{T^2}{T_0^2}\Theta(T_0-T)\Theta(T),
\end{equation}
with $\omega_1\ll \omega_0$. This modification allows burning at all
temperatures, what is more physically correct. The condition
$\omega_1\ll \omega_0$ imply that the model correction does not influence
on the stationary flame and the previous theory. So when $\omega<\omega_0$ the
flame spreads with a
constant velocity. When $\omega>\omega_0$ the evolution changes:
firstly according to previous
calculations front decays and smoothes in the ``thermoconductivity''
regime, but after it a slow burning in region $T<T_0$ raises the temperature to
the critical value and the flame blaze up again. An example of such propagation {\em for $\omega_0=8$} is
shown in Fig.~\ref{fig:Tt_80}, whereas the front position versus time is shown in
Fig.~\ref{fig:omega8_mod}.

This evolution is called front pulsations. Such a regime is a little
bit different than ``classical'' pulsations: $v_{\rm fr}=v_0+v_1\,\sin at$.
Here the flame stops, blaze up and stops again. It moves by jerks. It is interesting
that intervals between the jerks are considerably longer than the value
$|2\pi/\mbox{Im}~p|$ determined in the linear theory presented above. This relationship as well as
the jerky motion of the front
take place even for $\omega_0$ that is only slightly higher than the threshold $\omega_0=6$.

\section{Discussion and Conclusions}

We have presented here a simple analytically-solvable model for flame propagation which exhibits a
different behavior depending on the dimensionless parameter {\sl Ze}: the effective 
Zeldovich number expressed through our dimensionless parameter $\omega_0$. The theory is
compared with numerical
simulations, and a good agreement is observed.  When {\sl Ze}~$<6$ both
analytic and numerical solutions exhibit a constant-velocity
front. Our analytical theory shows that
the traveling flame front becomes unstable at {\sl Ze}~$>6$, whereas our numerical
simulations show that the front is destroyed and passes to jerk--like pulsations.

We believe that, in the case of pulsating instability in a real
flame, the characteristic behavior will be the same as in this model,
but the critical {\sl Ze} can be different. Therefore,  the proposed model can mimic the behavior
of numerical methods, and serve as a test for them.
Our present work cannot say anything about the existence of such pulsations in real white dwarfs.
However, if such a pulsating jerk-like
regime of slow flame propagation indeed takes place in white dwarfs,
then it could be able to trigger the transition to detonation.

We gratefully acknowledge extensive discussions with S.~I.~Blinnikov
and B.~Meerson.
The work is supported partly by grants RFBR 10-02-00249-a, RFBR 11-02-00441-a, ``Dynasty''
foundation, SCOPES project No.~IZ73Z0-128180/1, by Federal Programm ``Scientific and
pedagogical specialists of innovation Russia'' contract number
02.740.11.0250, and by the contract No.~02.740.11.5158 of the Ministry of Education and Science of the Russian Federation.

\bibliographystyle{apsrev}
\bibliography{toymodel}

\end{document}